\documentclass{iaus}
\usepackage{graphicx}

\title[Evolution and appearance of Be stars in SMC clusters] %% give here short title %%
{Evolution and appearance of Be stars in SMC clusters}

\author[C. Martayan et al.]   %% give here short author list %%
{C. Martayan$^{1,2}$, D. Baade$^3$, Y. Fr\'emat$^4$ \and J. Zorec$^5$}

\affiliation{$^1$ European Organisation for Astronomical Research in the Southern 
Hemisphere, Alonso de Cordova 3107, Vitacura, Casilla 19001, Santiago 19, Chile \\ email: {\tt cmartaya@eso.org} \\[\affilskip]
$^2$ GEPI, Observatoire de Paris, CNRS, Universit\'e Paris Diderot, 
5 place Jules Janssen, 92195 Meudon Cedex, France \\
$^3$ European Organisation for Astronomical Research in the Southern 
Hemisphere, Karl-Schwarzschild-Str.\ 2, 85748 Garching b.\ M\"unchen, Germany \\
$^4$ Royal Observatory of Belgium, 3 avenue circulaire, 1180 Brussels, Belgium\\
$^5$ Institut d'Astrophysique de Paris, UMR7095, CNRS, Universit\'e Marie \& Pierre Curie, 
98bis Boulevard Arago 75014 Paris, France
}

\pubyear{2009}
\volume{266}  %% insert here IAU Symposium No.
\pagerange{119--126}
\jname{Star Clusters}
\editors{Richard de Grijs \& Jacques R. D. L\'epine, eds.}
\begin{document}

\maketitle

\begin{abstract}

Star clusters are privileged laboratories for studying the evolution of massive
stars (OB stars).  One particularly interesting question concerns the phases,
during which the classical Be stars occur, which unlike HAe/Be stars, are not
pre-main sequence objects, nor supergiants.  Rather, they are extremely rapidly rotating B-type
stars with a circumstellar decretion disk formed by episodic ejections of matter
from the central star. To study the impact of mass, metallicity, and age on the
Be phase, we observed SMC open clusters with two different techniques:  1) with
the ESO-WFI in its slitless mode, which allowed us to find the brighter Be and
other emission-line stars in 84 SMC open clusters  2) with the VLT-FLAMES
multi-fiber spectrograph in order to determine accurately the evolutionary
phases of Be stars in the Be-star rich SMC open cluster NGC 330.  Based on a
comparison to the Milky Way, a model of Be stellar evolution / appearance as a
function of metallicity and mass / spectral type is developed, involving the
fractional critical rotation rate as a key parameter.

\end{abstract}

\firstsection % if your document starts with a section,
\section{Observations, spectral analysis}

We performed an Halpha survey with the ESO-WFI (\cite[Baade et al. 1999]{Baade
et al. 1999}) in its slitless mode in the central parts of the SMC.  Three
million low-resolution spectra were obtained. They were extracted from the
images with {\it Sextractor} (\cite[Bertin \& Arnouts 1996]{Bertin et Arnouts
1996}) and we developed {\it Album} (\cite[Martayan et al. 2009]{Martayan et al.
2009})to identify the spectra
presenting H$\alpha$-emission. After astrometric
calibration, the extracted sources were cross-matched with the OGLE-II 
photometric catalogue (\cite[Udalski et al. 1998]{uu1998}).  Within the area
observed, results for 84 SMC open clusters are presented. In particular, the
results are compared with the ones from \cite[McSwain \& Gies (2005)]{McSwain et
Gies 2005} in Galactic open clusters. The ratios of clusters with and without 
of classical Be stars are compared w.r.t.\ metallicity (Z), spectral-type, and
age.  Observations were also carried out with the VLT-FLAMES (\cite[Pasquini et al. 2002]{Pasquini et al.
2002}) of the Be-star rich SMC open cluster NGC330 and its
vicinity. LR2 spectra (395-455nm, R=6400, H$\epsilon$, H$\delta$, H$\gamma$,
HeI447.1nm, MgII448.1nm) and LR6 spectra  (650-770nm, R=8600, H$\alpha$) were
aquired between October 2003 and September 2004. For each star,  the fundamental
parameters were determined with GIRFIT (\cite[Fr\'emat et al. 2006]{Fremat et
al. 2006}), taking account of fast-rotation effects with FASTROT (\cite[Fr\'emat
et al. 2005]{Fremat et al. 2005}).  In particular, rotational velocities were
determined as well as the statistical 
fractional angular breakup velocity for Be stars of different ages (see \cite[Martayan et al. 2007]{Martayan et al.
2007}).

\section{Observed metallicity effect on the relative frequency of Be stars} 
 
\begin{figure}[b]
% \vspace*{-2.0 cm}
\begin{center}
 \includegraphics[width=2.5in, height=4.0in,  angle=-90]{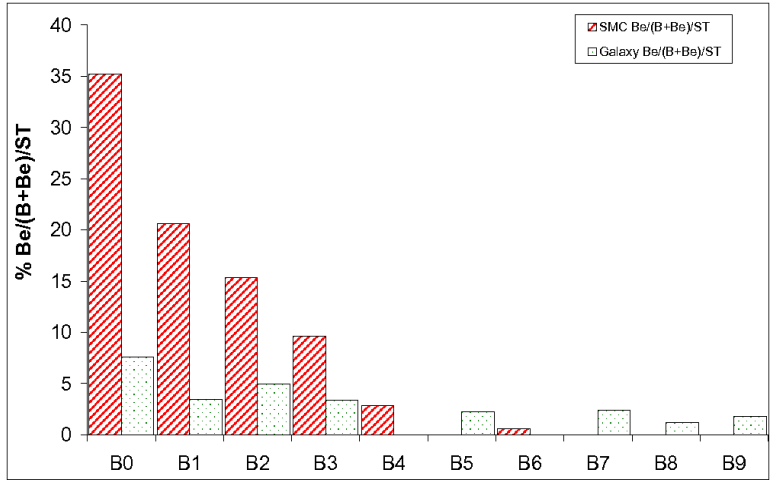} 
% \vspace*{-1.0 cm}
 \caption{The relative abundances of Be stars in the SMC (hatched bars) and the Galaxy (dot-filled bars) as a function of  spectral type.}
 \label{fig1}
\end{center}
\end{figure}

The metallicity of the SMC is significantly lower (0.001$<$Z$<$0.009, 
\cite[Cioni et al. (2006)]{Cioni  et al. (2006)}) than the one of the Galaxy
(Z=0.020).  Fig.~\ref{fig1} shows the fraction of Be stars to all B-type stars
by spectral type, separately for SMC and MW.  It clearly indicates that
early-type Be stars are 3-5 times as frequent in the SMC as in the MW (beyond
B2-B3, the SMC sample is incomplete). Note that the age ranges are comparable. 
The large difference in frequency is obviously most easily attributed to the
difference in metallicity.  

%The maximum of Be stars among all Be stars of all spectral-types is reached at spectral-type B2 in
%the SMC and in the MW. No effect of metallicity is seen here however the WFI study is not complete for %spectral-types
%later than B3.

\section{Appearance of Be stars in open clusters as a function of age}
 
\begin{figure}[b]
% \vspace*{-2.0 cm}
\begin{center}
 \includegraphics[width=2.5in, height=4.0in, angle=-90]{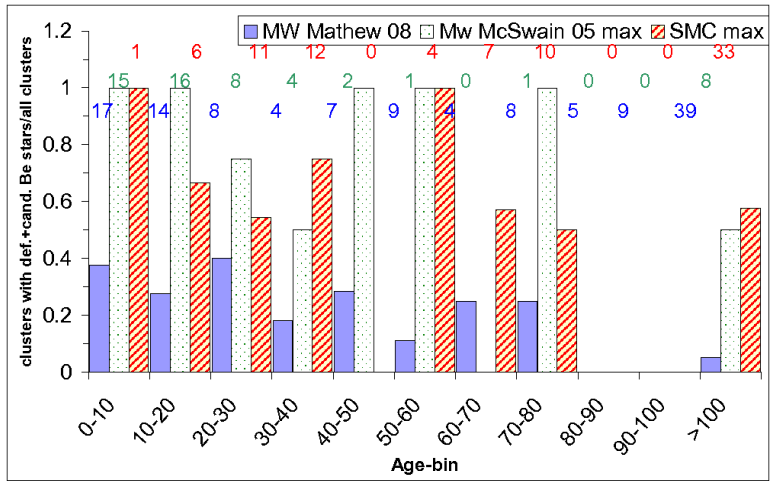} 
% \vspace*{-1.0 cm}
 \caption{Fraction of open clusters with Be-star members in the SMC (hatched bars) and the MW (dot-filled bars: data from
\cite[McSwain \& Gies (2005)]{McSwain et Gies 2005}, filled bars: data from \cite[Mathew et al. (2008)]{Mathew et al.
2008}) as a function of cluster age.  The number of clusters studied is given above each bar.}
\label{fig2}
\end{center}
\end{figure}

Fig.~\ref{fig2} displays the fraction of open clusters with Be-stars in SMC and
MW vs.\ age.  It appears that Be stars are preferrentially hosted by young open
clusters.  A first decrease in the fraction of open clusters with Be stars is
seen around  30-40 Myears (corresponding to early-type Be stars reaching the
TAMS) is followed by an increase of open clusters with Be stars.  The terminal
decrease in older open clusters is due to also the late-type Be stars arriving
on the TAMS.

\section{Metallicity-dependent evolution of rotational velocities}
 
To explain the overall shape of Fig.\ 2, the temporal evolution of the
fractional angular velocity ($\frac{\Omega}{\Omega_{c}}$) of Be stars can be
reconstructed from observed rotational velocities  (here we use VLT-FLAMES
spectra) and theoretical evolutionary tracks from  \cite[Maeder \& Meynet
(2001)]{MaederMeynet 2001}. From our SMC data and results obtained by 
\cite[Zorec et al. (2005)]{Zorec et al. (2005)} in the MW, Be stars have a
minimal   $\frac{\Omega}{\Omega_{c}}$ of 0.7.  If/when B-type stars rotate more
slowly, Be symptoms do not seem to develop.

The evolution of the $\frac{\Omega}{\Omega_{c}}$ depends on stellar mass and
metallicity  (see Fig.\ 5 from \cite[Martayan et al. (2007)]{Martayan et al.
2007}), which holds the key to the understanding of the distribution seen in
Fig.\ 2:\\
\begin{list}{\itemsep=0mm \topsep=0mm}
\item
In massive MW Be stars, the $\frac{\Omega}{\Omega_{c}}$ sufficient for the
development of Be star symptoms at the beginning of the Main Sequence.  Due to
wind-driven mass loss the angular momentum decreases until
$\frac{\Omega}{\Omega_{c}}$  has dropped below the threshold of 0.7 and the
initial Be-star  appearance will be lost after $\sim$5-10 Myears. This can
explain the small decrease observed in the MW for open clusters with Be stars as
seen in the data of \cite[Mathew et al. (2008)]{Mathew et al. 2008}. 

\item
Intermediate-mass Be stars seem to retain a sufficently high level of
$\frac{\Omega}{\Omega_{c}}$ during their entire Main Sequence evolution to
preserve their appearance as Be stars without interruption.   They Be-stars
attributes will eventually diseappear once the TAMS is reached at $\sim$
40Myears, which can explain the lower fraction of  Galactic open clusters with
Be stars of this age. 

\item
Rapidly rotating low-mass B stars can appear as Be stars at the beginning of the
Main Sequence but will lose this status quickly by a reorganization of the
internal angular momentum.  Later, the standard evolution of 
$\frac{\Omega}{\Omega_{c}}$ with time (increase of the radius with small angular
momentum loss), will let them reach an  $\frac{\Omega}{\Omega_{c}}$ above 0.7 so
that at an age of 40-50 Myears they are  again recognizeable as Be stars.  This
fact can explain the second increase of the number of open clusters with Be
stars at an age 40 Myears.  The low-mass Be stars will then begin to reach the
TAMS and the number of open clusters with Be stars decreases with age.

\item
In the SMC,  the evolution of $\frac{\Omega}{\Omega_{c}}$ of intermediate- and
low-mass Be stars is similar to the MW case because in this mass range mass-loss
is too small to significantly alter the angular momentum or its internal
distribution.  This explains the evolution of open clusters with Be stars and
ages between 30 and 100 Myears.  Still older open clusters that host Be stars
probably had multiple star-formation episods or the Be stars may be blue
stragglers. Note that, for the MW, \cite[McSwain \& Gies (2005)]{McSwain et Gies
2005} suspect that 75\% of the Be stars are binaries. 

\item
The evolution of more massive Be stars is different in the SMC than in the MW. 
Because of the lower SMC metallicity, mass and angular momentum loss are lower
as well, and $\frac{\Omega}{\Omega_{c}}$ will  increase with the time as in
low-mass Be stars.  They will reach the 70\% limit at $\sim$4-5 Myears and then
again appear as Be stars, in contrast massive Be stars in the MW.  Therefore, in
the SMC very young open clusters are expected to contain Be stars.  
\end{list}

However, for the observational verification of this difference in the fraction
of Be stars between very young open clusters in MW and SMC a larger sample of
open clusters younger than 10 Myears is needed.  For such a comparison, one must
careful discriminate between classical Be stars and Herbig Ae/Be stars, which
are still on the pre-MS  and owe their emission lines to their parental
accretion disk.  A method for this separation  is described in \cite[Martayan et
al. (2008)]{Martayan et al. (2008)}. 

The inferred long-term transitions between Be and B phases will not probably be
abrupt, and one may speculate whether the much faster and repetitive Be to B and
B to Be transformations observed in many Be stars are part of the long-term
transitions.  If so, one would expect that Be stars with stable and Be stars
with more volatile circumstellar disks should not show gross differences
otherwise.  In fact, this is what \cite[McSwain et al. (2009)]{McSwain et al. 2009},
reported recently.  
\section{Conclusions}
 
We conducted a study of open clusters with Be stars in two data sets:
\begin{itemize}
\item
a low-resolution H$\alpha$ survey of 84 SMC open clusters.  It indicates that Be
stars are more abundant in the SMC than in the MW.  The fraction of open
clusters with Be stars reaches a local minimum at ages around  30-40 Myears and
finally declines for clusters older than 70-80 Myears. 
\item
Fundamental stellar parameters determined from medium-resolution spectra of B
and Be stars in the SMC open cluster NGC330 and its vicinity allowed us to study
the evolution of the fractional critical angular velocity,
$\frac{\Omega}{\Omega_{c}}$, of Be stars.  Its dependency on stellar mass and
metallicity can explain the distribution with age of the fraction of open
clusters with Be stars and the differences between Galaxy and SMC.
\end{itemize}
All data and a more comprehensive discussion are provided in \cite[Martayan et
al. (2009)]{Martayan et al. 2009}.

\end{document}